\documentclass[letterpaper,aps,pre,twocolumn,showpacs,preprintnumbers,superscriptaddress]{revtex4}

\usepackage{graphicx}% Include figure files
\usepackage{dcolumn}% Align table columns on decimal point
\usepackage{bm}% bold math

\begin{document}

\title{Dynamic effects induced by renormalization in anisotropic pattern forming systems}

\author{Adrian Keller}
\affiliation{Institute of Ion Beam Physics and Materials Research,
Helmholtz-Zentrum Dresden-Rossendorf, P.O. Box 510119, 01314
Dresden, Germany}
\affiliation{Interdisciplinary Nanoscience Center (iNANO), Aarhus University, Ny Munkegade, 8000 Aarhus C,
Denmark}
\author{Matteo Nicoli}
\affiliation{Physique de la Mati\`ere Condens\'ee, \'Ecole Polytechnique - CNRS,
91128 Palaiseau, France}
\author{Stefan Facsko}
\affiliation{Institute of Ion Beam Physics and Materials Research,
Helmholtz-Zentrum Dresden-Rossendorf, P.O. Box 510119, 01314
Dresden, Germany}
\author{Rodolfo Cuerno}
\affiliation{Departamento de
Matem\'{a}ticas and Grupo Interdisciplinar de Sistemas Complejos
(GISC), Universidad Carlos III de Madrid, Avenida de la
Universidad 30, E-28911 Legan\'{e}s, Spain}

\date{\today}

\begin{abstract}
The dynamics of patterns in large two-dimensional domains remains a challenge in non-equilibrium phenomena. Often it is addressed through mild extensions of one-dimensional equations. We show that full 2D generalizations of the latter can lead to unexpected dynamical behavior. As an example we consider the anisotropic Kuramoto-Sivashinsky equation, that is a generic model of anisotropic pattern forming systems and has been derived in different instances of  thin film dynamics. A rotation of a ripple pattern by $90^{\circ}$ occurs in the system evolution when nonlinearities are strongly suppressed along one direction. This effect originates in non-linear parameter renormalization at different rates in the two system dimensions, showing a dynamical interplay between scale invariance and wavelength selection. Potential experimental realizations of this phenomenon are identified.

\end{abstract}


\mbox{Published as Phys.\ Rev.\ E {\bf 84}, 015202(R) (2011); {\bf 85}, 029905(E) (2012).}

\maketitle

The self-organized formation of patterns in non-equilibrium systems is a fascinating topic that has focused a large attention in the last decades. Examples range from galaxy formation to sandy dunes, to nanostructures \cite{cross_book}.
While regular patterns like stripes and hexagons that are characterized by a single length-scale $\ell$ are well understood when the lateral system size $L$ is comparable to $\ell$, their dynamics becomes much more complex in the large domain limit $L \gg \ell$. Indeed, intricate structures ensue, like spatiotemporal chaos, spiral waves or quasiperiodic patterns \cite{cross}. Another source of complexity derives from dimensionality. While a unified description of one-dimensional patterns is available through the Ginzburg-Landau equation \cite{cross_book}, this is not the case for two-dimensional systems, strongly anisotropic problems providing particularly challenging cases.

Many systems depending of two space variables are studied within the context of 3D localized structures \cite{toh89}, like vortices in plasmas \cite{kuznetsov86}, or solitary waves in fluids \cite{chang02,demekhin10}.
The dynamic equations considered are frequently mild extensions of 1D equations in which only specific terms are turned into 2D operators. This allows to probe the behavior of a given localized structure when the spatial dimension is increased.
In other cases, the equation derives from first principles, as when studying anisotropic surface tension and kinetics in solidification systems \cite{davis_book}. A prototype model appearing in all these studies is the anisotropic Kuramoto-Sivashinsky (aKS) equation
\begin{eqnarray}
\partial_t h&=& %v_0 +\gamma \partial_x h
\nu_x \partial_x^2 h+\nu_y \partial_y^2 h +\frac{\lambda_x}{2} \left (\partial_x h\right)^2
+ \frac{\lambda_y}{2} \left (\partial_y h\right )^2
\nonumber\\
& & - {\cal K}_{x} \partial_x^4 h - 2 {\cal K}_{xy} \partial_x^2\partial_y^2 h - {\cal K}_{y} \partial_y^4 h +\eta.
\label{KSE}
\end{eqnarray}
For the sake of definiteness, we will keep in mind a physical picture in which $h(x,y,t)$ is interpreted as the height of a surface
above point $(x,y)$ on a reference plane at time $t$. Indeed, particular instances of Eq.\ (\ref{KSE}) have been derived in various contexts
of thin film dynamics like surface nanopatterning by ion-beam erosion \cite{cuerno95}, epitaxial growth \cite{villain91,wolf91,rost_prl} or solidification from a melt \cite{golovin98}. In Eq.\ (\ref{KSE}),
the morphological instability leading to pattern formation is implemented by the coefficients $\nu_{x,y}$, at least one of them being {\em negative}. Terms with coefficients ${\cal K}_j$ provide dissipation at the smallest scales, while the nonlinearities proportional to $\lambda_{x,y}$ stabilize the system. We have incorporated a Gaussian, zero-average, uncorrelated noise $\eta(x,y,t)$ as a means to explore the aKS equation in a large domain. Indeed, for $L \gg \ell$, the deterministic KS equation is well known to display spatio-temporal chaos and a steady state with strong height fluctuations \cite{krug97,pradas11}. Introduction of noise helps to elucidate these for times after onset of the morphological instability \cite{cuerno_kpz_prl,ueno,nicoli10}, while it is not essential for the occurrence of the morphological transition we are studying in this paper.

Regarding Eq.\ (\ref{KSE}) as a model of 3D localized structures, e.g.\ the description of solitary fluid waves moving down an inclined plane corresponds to $\nu_y=\lambda_y=0$, ${\cal K}_x = {\cal K}_y = {\cal K}_{xy}$ \cite{petviashvili81,demekhin10}, while in the solidification system $\nu_y=-\nu_x$, $\lambda_y={\cal K}_y={\cal K}_{xy}=0$ \cite{davis_book}.
Thus Eq.\ (\ref{KSE}) is tailored to preserving the quasi one-dimensional features of specific solutions. In this work, we show that small deviations from fine tuned conditions such as the latter are able to induce dynamic effects in the system that unavoidably require a full two-dimensional description. Specifically, a rippled pattern appearing at short times along one of the system directions rotates by $90^{\circ}$ during the evolution, leading at longer times to ripples oriented in the perpendicular direction. This dynamic transition occurs as a result of the different rates at which fluctuations renormalize due to the inhomogeneous strengths of nonlinear effects along the two space dimensions.

We start by exploring numerically the behavior of Eq.\ (\ref{KSE}).
It is convenient to first bring it to dimensionless form. As the properties to be studied are not
conditioned by the anisotropy of the dissipative terms, we will
restrict ourselves to the case in which ${\cal K}_x = {\cal K}_y = {\cal K}_{xy}
\equiv {\cal K}$. Then, defining rescaled coordinates $t'=-(|\nu_x|\nu_x/{\cal K})t$, $\textbf{x}' =
(|\nu_x|/{\cal K})^{1/2}\textbf{x}$, $h'= -(\lambda_x/2\nu_x)h$
leads to
\begin{eqnarray}\label{min_aKSE}
\hspace*{-15pt} \partial_t h&\hspace*{-3pt} =& \hspace*{-3pt}-\partial_x^2 h-a_{\nu} \partial_y^2 h+\left
(\partial_x h\right )^2 %\nonumber\\
%& &{}
+a_{\lambda}\left (\partial_y h\right )^2 \hspace*{-4pt}-\nabla^4h+\xi ,
\end{eqnarray}
where primes are dropped, $\xi$ is a noise term with a rescaled variance,
and the ratios $a_{\nu}=\nu_y/\nu_x$ and $a_{\lambda}=\lambda_y/\lambda_x$
control the linear and the nonlinear anisotropies, respectively.
We perform numerical simulations of Eq.\ (\ref{min_aKSE}) for flat initial conditions
by using both a finite difference scheme with periodic boundary conditions \cite{keller_book}, and
alternatively a pseudospectral scheme \cite{giada02,giada02b}.

\begin{figure}[t!]
\includegraphics[angle=-90,width=0.5\textwidth,clip=]{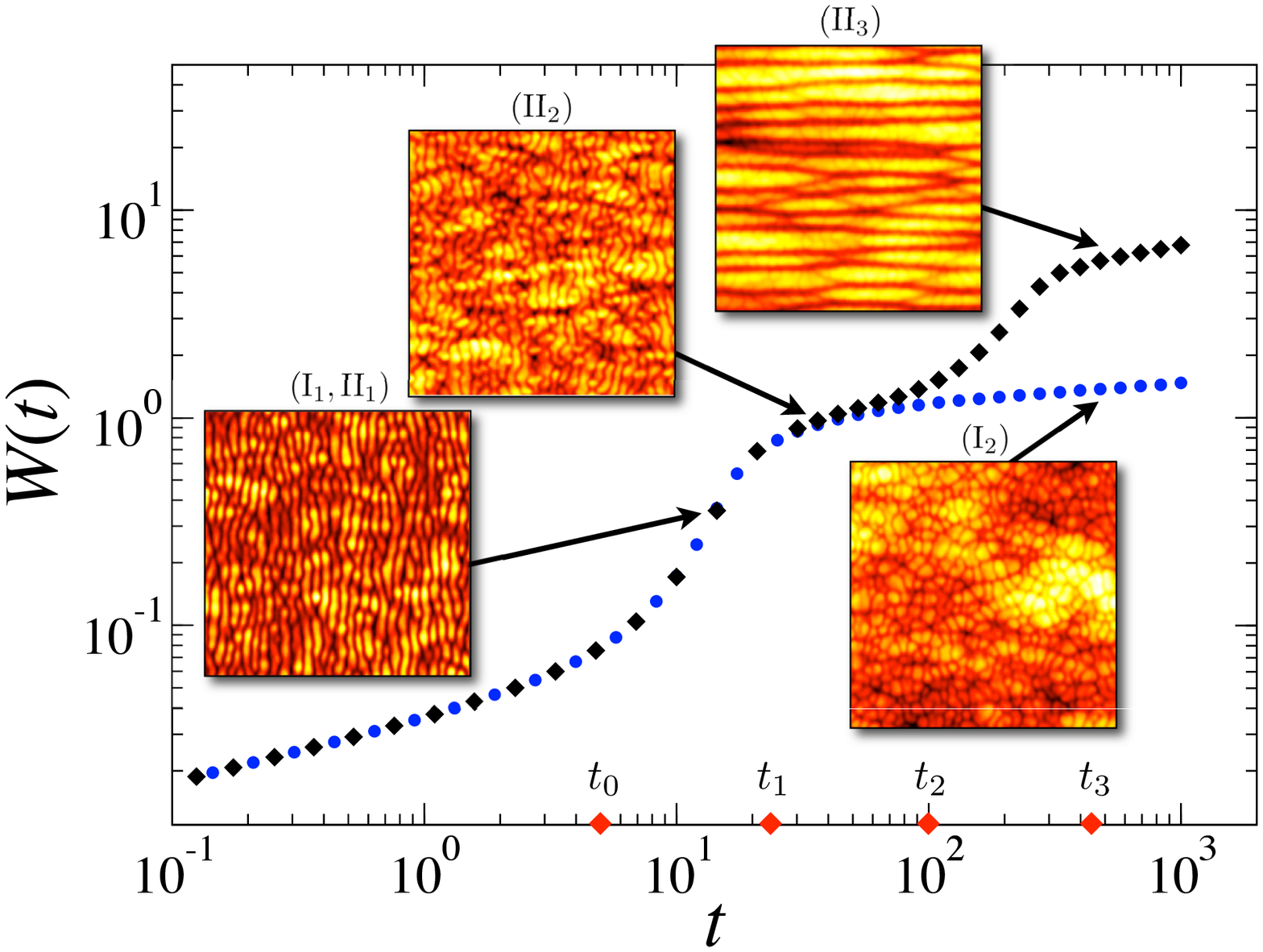}
\caption{\label{ev_aKS} Surface roughness vs time for conditions I
(circles) and II (diamonds). Simulations of Eq.\ (\ref{min_aKSE}) are performed with $a_{\nu}=0.1$,
$D=10^{-2}$, $\Delta t = 5\times 10^{-3}$, $L=512$, $\Delta x = 1$, and $a_\lambda = 0.5$ (I), $0.025$ (II).
$W(t)$ is averaged over $100$ noise realizations. Top views of the surface morphology (size $L/2$; see
\cite{epaps} for larger sizes) are shown for conditions I and II at times as indicated by arrows.
Times $t_j$ in the main text appear as diamonds on the horizontal axis. All units are arbitrary.
}
\end{figure}

Fig.\ \ref{ev_aKS} shows the time evolution of the surface roughness
$W^2(t) = \langle (1/L^2) \sum_{\mathbf{r}} (h_{\mathbf{r}}(t) - \bar{h}(t))^2 \rangle$, where brackets
denote average over noise realizations and bar denotes space average. Two
different parameter conditions are considered for $a_{\nu} < 1$ \cite{epaps}.
In the first one (I), the nonlinear couplings are comparable in the two directions
$x$ and $y$ ($a_{\lambda}=0.5$), while the second one (II) is representative of conditions
in which $\lambda_y$ is strongly suppressed ($0 < a_{\lambda} \lesssim 0.1$).
Values of time at which $W(t)$ changes behavior significantly are marked by showing
the corresponding surface morphologies. For both conditions, $a_{\nu}<1$ induces
a linear instability leading to formation of a ripple structure with crests oriented parallel to the $y$ axis.
This takes place at time $t_0$ in Fig.\ \ref{ev_aKS}, at which $W(t)$
grows exponentially. The corresponding morphologies are statistically
indistinguishable for both conditions, so that a single common snapshot (I$_1$,II$_1$) is
shown in Fig.\ \ref{ev_aKS}. For later times, this regime is followed
by non-linear stabilization inducing at time $t_1$ a slower, power-law growth rate for $W(t)$.
Morphologically, this type of growth is characterized by a progressive blurring of the early time pattern
(morphology I$_2$) and the dominance of height fluctuations associated with kinetic roughening \cite{krug97,park}.
For condition II, noise effects also dominate for time $t_1<t<t_2$ (morphology II$_2$). At time $t=t_2$
there is a second rapid increase in $W(t)$ that morphologically corresponds to the formation
of a new ripple structure, but now with crests parallel to the $x$ axis. The wave-length of
this new pattern is larger than that of the initial one, see morphology II$_3$. Again, nonlinear
effects stabilize growth in amplitude for times $t>t_3$, so that $W(t)$ displays kinetic roughening properties similar
to those occurring for condition I at long times \cite{park}.

In the simulations, $a_{\lambda}$ is always positive, so that
the known cancellation modes of the aKS equation \cite{rost_prl} are
not solutions of Eq.\ (\ref{min_aKSE}).
The dynamic behavior seen so far can be interpreted using results for the
noisy 1D KS equation. Thus, as already argued for by Yakhot \cite{yakhot81} in the deterministic limit,
the noisy KS (nKS) equation is known to undergo a renormalization process through which the
negative coefficient of the second order linear term becomes effectively positive (and therefore
stable) at sufficiently large space and time scales \cite{cuerno_kpz_pre,ueno}. This effect is induced by the nonlinearity
and allows for its eventual control of the scaling behavior at the stationary state, that is in the Kardar-Parisi-Zhang
(KPZ) universality class \cite{krug97}. Analogous behavior has been recently shown to occur in the 2D isotropic nKS case
\cite{nicoli10}. Morphologically, the linear pattern forming instability occurring at short times
is followed by a disordered height morphology showing kinetic roughening properties of the KPZ class asymptotically.

In the case of the aKS equation (\ref{min_aKSE}) with small $a_{\lambda}$, this
renormalization can be expected in the $x$ direction, resulting in
a stabilizing value $\nu^*_x>0$. In the $y$ direction, however, the
nonlinearity is so weak that the transition to the nonlinear
regime is strongly delayed. Therefore, at times $t\simeq t_2$ when
$\nu^*_x$ has already renormalized to a positive value, the
corresponding $\nu^*_y$ coefficient has not yet, and remains
negative. Then, a new linear instability in the $y$ direction causes the formation
of a ripple pattern that appears rotated by 90$^{\circ}$ with respect to
the early time pattern. These ripples then grow exponentially in time until
the values of the slopes along the $y$ direction become so large that
the corresponding nonlinearity in Eq.\ (\ref{min_aKSE}) takes over
and the new ripple amplitude is stabilized. One can estimate
the value of the time $t_3$ at which this happens to be \cite{park}
\begin{equation}\label{an_t_c}
t_{3}\propto
% \frac{1}{a_{\nu}}\ln\left(\frac{a_{\nu}}{a_{\lambda}}\right).
a_{\nu}^{-2} \ln\left(a_{\nu} / a_{\lambda}\right) .
\end{equation}
This expression arises from the behavior of the surface roughness following the linearized equation (\ref{min_aKSE}) up to times  $t \leq t_3$ as $W(t_3) \sim \exp(a_{\nu} t_3/ \ell_{y}^2)$, the assumption of a similar scaling for surface height and roughness, $h \sim W$, and a balance between the linear and non-linear terms in Eq.\ (\ref{min_aKSE}) precisely at time $t_3$, namely $a_{\nu} \partial_y^2 h \sim a_{\lambda} (\partial_y h)^2$, implying $h \sim a_{\nu}/a_{\lambda}$.
The ``linear'' wavelength of the rotated pattern should moreover be given by
$\ell_y=2\pi(2/a_{\nu})^{1/2}$.

\begin{figure}[t!]
\includegraphics[angle=-90,width=0.5\textwidth,clip=]{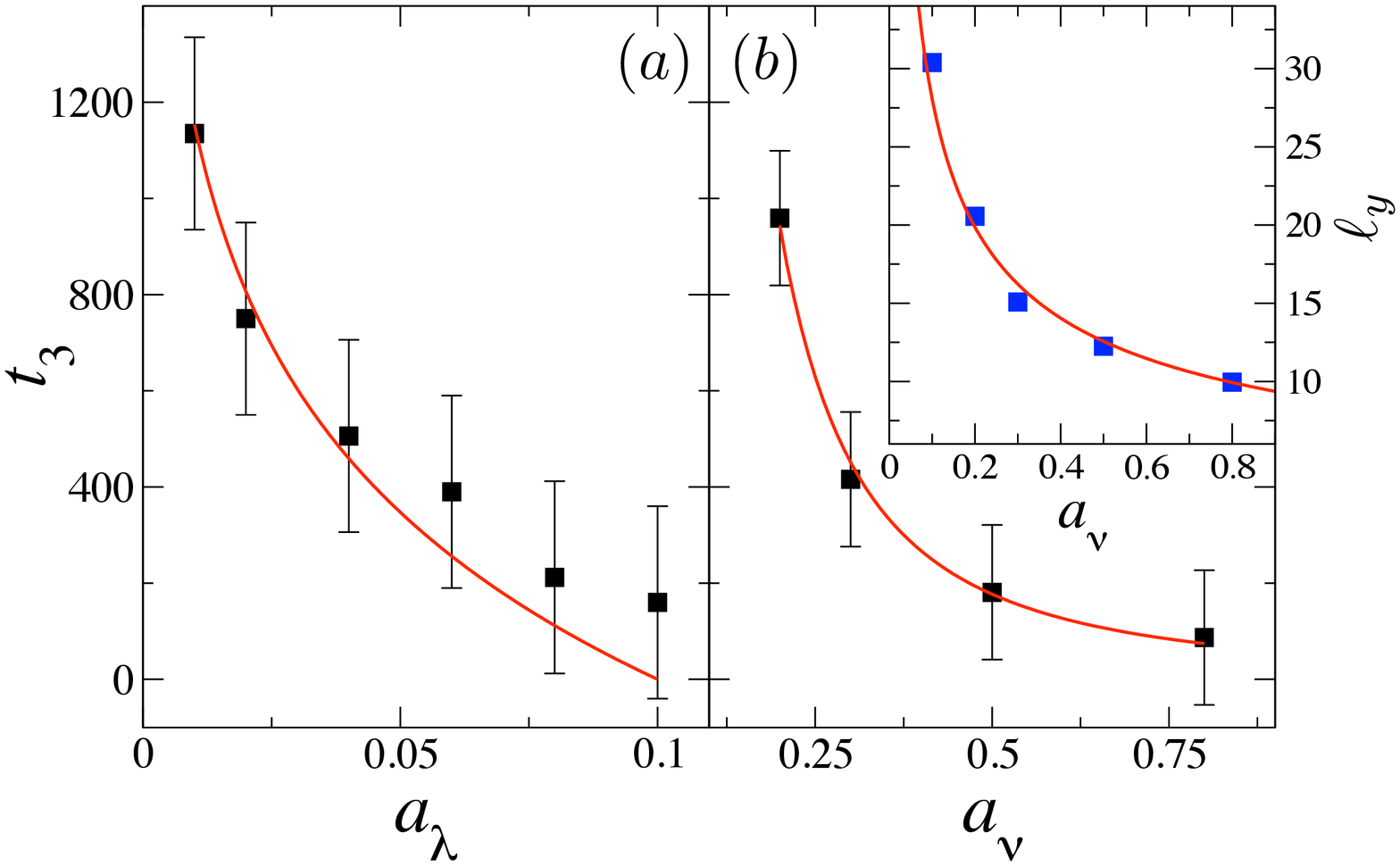}
\caption{\label{tc2} Transition time $t_{3}$ vs (a) $a_{\lambda}$
for $a_{\nu}=0.1$ and (b) $a_{\nu}$ for $a_{\lambda}=10^{-3}$.
Solid lines represent fits according to Eq.\ (\ref{an_t_c}) with
a prefactor as a fitting parameter. The inset in (b) shows the
wavelength $\ell_y$ of the rotated pattern as a function of
$a_{\nu}$, as determined from the simulations for $a_{\lambda}=10^{-3}$
(blue squares; error bars are smaller than the symbol size) and
calculated within a linear approximation %as $\ell_y=2\pi(2/a_{\nu})^{1/2}$
(red line). All units are arbitrary.}
\end{figure}

Quantitative comparisons between the simulated and the analytical
dependence of $t_{3}$ on $a_{\lambda}$ and $a_{\nu}$ are shown in
Figs.\ \ref{tc2}(a) and \ref{tc2}(b), respectively. Also,
$\ell_y$ is shown in the inset of Fig.\ \ref{tc2}(b) as a function of $a_{\nu}$ as obtained from simulations,
and as calculated from the linear approximation.
In all cases, values from simulations agree well with analytical
estimates, supporting our interpretation of the dynamic morphological
transition within the framework of parameter renormalization.

\begin{figure}[t!]
\includegraphics[angle=-90,width=0.475\textwidth,clip=]{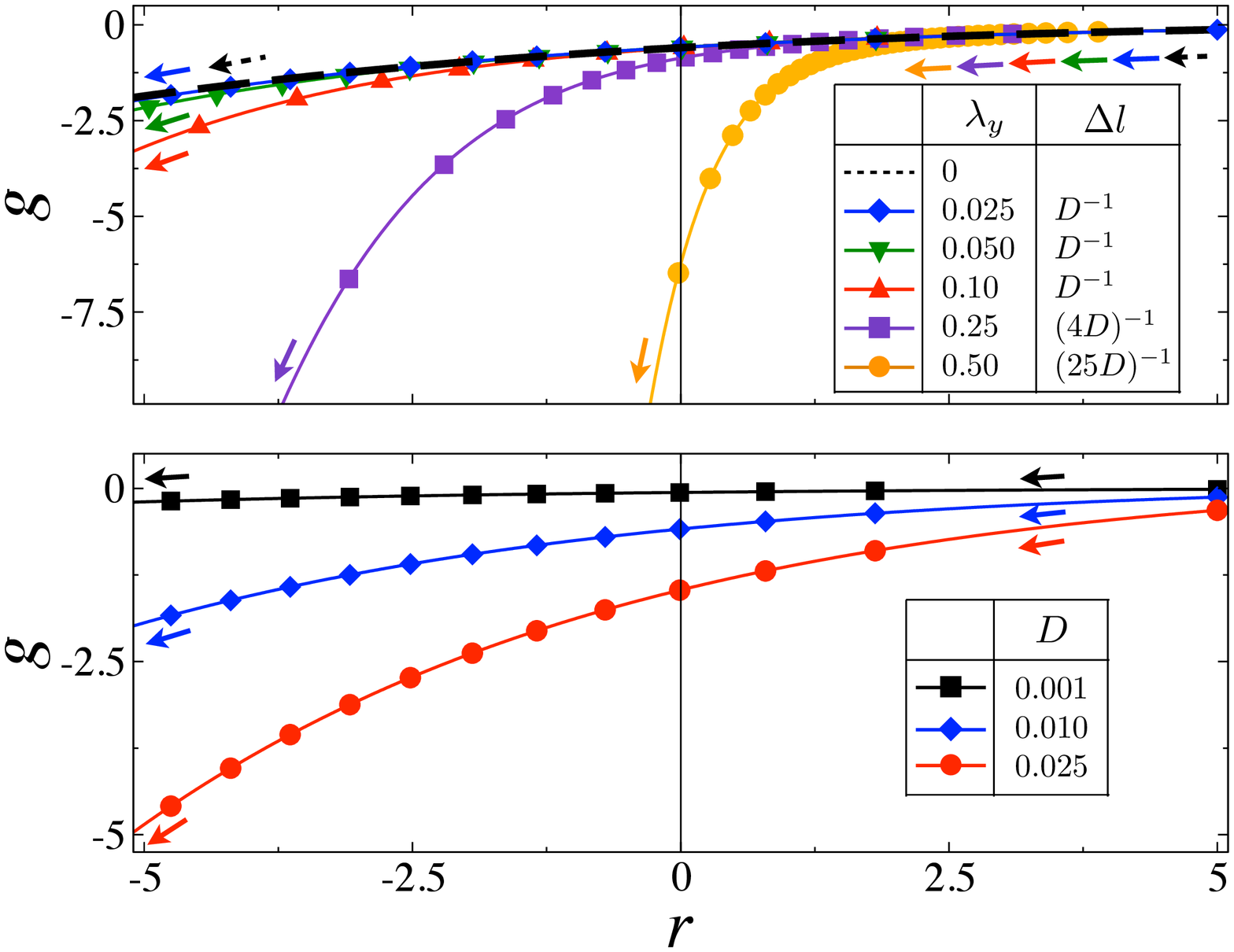}
\caption{\label{flow} DRG flow of the couplings $r$ and $g$. The numerical
integration of equations (\ref{aks:red}) has been carried out using $\nu_x = -1$,
$\nu_y = -0.2$, $\lambda_x = 1$, ${\cal K}_x = {\cal K}_y = {\cal K}_{xy} = 1$, and values of
$\lambda_y$ and $D$ as shown in the legends. Upper panel: fixed $D=10^{-2}$ and changing $\lambda_y$. Note that
$g$ renormalizes even for $\lambda_y = 0$. Markers display values of $(r(l),g(l))$
separated by $\Delta l$ as shown in the legend, while arrows show the direction of the RG flow.
Lower panel: for fixed $\lambda_y = 0.025$, and different values of $D$ with $\Delta l = 1/D$.
All units are arbitrary.}
\end{figure}

Further progress is possible by studying Eq.\ (\ref{KSE}) within a one loop Dynamical Renormalization Group (DRG)
approach. After its application to fluctuating hydrodynamics \cite{mccomb91}, this method has recently shown a
large explanatory power in related contexts, like a multiscale descriptions of fluctuating interfaces
\cite{haselwandter07} or morphological instabilities mediated
by non-local interactions \cite{nicoli09}.
Following the standard approach \cite{mccomb91}, we arrive at the following RG parameter flow \cite{epaps},
\begin{equation}
\label{aks:red}
\frac{d r}{d l} = r \left(
	\frac{\Sigma_{\nu_y}}{\nu_y}-\frac{\Sigma_{\nu_x}}{r \nu_y}\right), \quad %\\[5pt]
\frac{d g}{d l} = g\left(
	3\frac{\Sigma_{\nu_y}}{\nu_y} +\frac{\Phi}{ D} \right) ,
%	\label{aks:red-g}
\end{equation}
where a coarse-graining of the height and noise fields has been performed in a fast mode shell in
wave-vector space $\mathbf{k}$ with $k \in [\Lambda (1-dl), \Lambda]$, where $\Lambda=1$ is a lattice cut-off.
Here, $r = \nu_x/\nu_y$, $g = \lambda_x^2 D/(\pi^2 \nu_y^3)$, with $D$ being the noise variance.
Eqs.\ (\ref{aks:red}) generalize the DRG analysis of the anisotropic KPZ equation in \cite{wolf91}.
The functions $\Sigma_{\nu_j}$ originate from propagator renormalization, whereas $\Phi$ arises in noise variance
renormalization \cite{epaps}. The flow (\ref{aks:red})
has been derived within the further assumption that, as expected \cite{cuerno_kpz_pre,ueno},
the parameters ${\cal K}_j$ become enslaved \cite{haken_book} to the slower parameter $\nu_y$ in Eq.\ (\ref{KSE}).
In Fig.\ \ref{flow} we show results from a numerical integration of (\ref{aks:red}).
Starting out from conditions for which $\nu_x$ and $\nu_y$ are both negative, parameter $r$ is seen
to cross the zero value meaning that, at the corresponding scale $l$, $\nu_x(l)$ has become
positive and stabilizing, while $\nu_y(l)$ remains negative. After further coarse-graining,
$g$ decreases to $-\infty$, signalling renormalization of $\nu_y$ towards positive values which
in turn requires crossing $\nu_y=0$ at the appropriate (large) scale. Note that,
the larger $\lambda_y$ is, the faster renormalization of $g$ takes place, as denoted by the
relative spacing among points evaluated at equally spaced positions $n \Delta l$ along the corresponding
flow trajectories.
Once both $\nu_j$ coefficients have renormalized to positive values, by analogy with the 1D and the isotropic 2D cases
for the nKS equation one expects the system to enter the anisotropic KPZ regime \cite{wolf91}. Within this picture,
a stationary state is expected at long times, that shows kinetic roughening. As in the noisy KS case
\cite{cuerno_kpz_prl,ueno}, this state is more efficiently reached when the noise variance is larger,
as seen in the lower panel of Fig.\ \ref{flow}, where $g$ renormalizes faster for
increasing $D$ values. Indeed, the slow growth of $W(t)$ at long times for condition II in Fig.\ \ref{ev_aKS}
signals the stabilization of the second ripple structure by non-linear effects. Note, this is also the case for condition I for which
no second ripple structure exists. Thus, bare values of the nonlinearities that are comparable to each other (condition I) lead to faster renormalization of $\nu_{x,y}$ to stable positive values and to a rough, disordered stationary state, as implied by the
upper panel in Fig.\ \ref{flow}.

The present dynamic morphological transition induced by renormalization effects can be intuitively understood by an argument that
employs knowledge of a particular solution of Eq.\ (\ref{KSE}) for the case in which $\lambda_y=0$. In this extreme
limit one can assume \cite{rost_prl} that $h(x,y,t)=H(y,t)$ does {\em not} depend explicitly on $x$ so that $\partial_x^n H=0$, even for
non-zero $\nu_x$ and $\lambda_x$.
Substituting this Ansatz into Eq.\ (\ref{KSE}), we get $\partial_t H = \nu_y \partial_y^2 H - {\cal K}_y \partial_y^4 H
+ \eta$, that is a linearized 1D noisy KS equation for $H$ because $\nu_y<0$. Therefore $H(y,t)$ is a ripple structure with crests along the $x$ direction. Moreover, linear stability analysis leads to a dependence of the wavelength of this ripple structure precisely as determined for $\ell_y$ in Fig.\ \ref{tc2} once rescaled coordinates are employed as in Eq.\ (\ref{min_aKSE}).
Note, however, that $H(y,t)$ {\em is not} a solution of Eq.\ (\ref{KSE}) for conditions II. Still, starting out from a flat initial surface, the term ${\cal N}_y = \lambda_y (\partial_y h)^2$ stays negligible until $t \simeq t_2$ at which the full solution $h(x,y,t) \approx H(y,t)$. Since the latter solution is itself morphologically unstable, large values of $\partial_y h \approx \partial_y H$ build up that make the term ${\cal N}_y$ no longer negligible. This introduces significant differences between $h$ and $H$ after $t> t_3$. A limitation of this argument is its neglect of dynamics up to $t=t_2$. At earlier times, a ripple structure exists in the perpendicular direction, for which it is $y$-derivatives, rather than $x$-derivatives, that are small (see e.g.\ morphology I$_1$, II$_1$ in Fig.\ \ref{ev_aKS}). For solution
$H(y,t)$ to become dynamically relevant, the initial ripple structure need to be washed out by fluctuations and nonlinearity as in condition I, which requires the renormalization process discussed above.

A dynamical transition reminiscent of the one considered here has been observed experimentally
in high-temperature surface nanopatterning by ion-beam sputtering (IBS) of Si(111) surfaces \cite{brown_prl,brown_prb}.
Specifically, at low fluence (equivalent to time for the fixed flux conditions employed), a ripple pattern with crests
perpendicular to the direction of the incident ion beam formed on the surface, with a wavelength $\ell_x \simeq 300 - 500$ nm.
At intermediate fluence, however, a different ripple pattern rotated by 90$^{\circ}$ overlayed the initial one, resulting in a
pattern of dot-like features. At even higher fluence, the initial pattern vanished and only the rotated pattern, with a significantly
larger wavelength $\ell_y > 500$ nm remained. The experimentally observed rotation of the ripple pattern by 90$^{\circ}$ does not agree
with the predicted angle for cancellation modes, which under these experimental conditions is expected to be $25^{\circ}$ \cite{brown_prl}. Hence, the observed ripple rotation is not related to the appearance of cancellation modes and must be of a different origin.
Given the striking similarities to the transition studied in our work, a strong nonlinear anisotropy with $a_{\lambda}\ll 1$ can be assumed for the experimental system. Transient morphologies of two-dimensional features similar to those observed in \cite{brown_prl,brown_prb}
can also be achieved in the simulations of the aKS equation by tuning the $a_{\nu}$ and $a_{\lambda}$ coefficients in such a
way so that the growth of the rotated ripples sets in before the initial pattern has fully vanished \cite{epaps}.
Although the physical picture leading to the aKS equation as a physical model for IBS has been recently contested
(see a review in \cite{cuerno11}), the high temperature condition employed in these experiments
can be expected to enhance surface transport. In such a case, an aKS-type equation
holds with modified coefficients \cite{cuerno11}, which may moreover account for the lack of cancellation modes in the experiments as compared with theoretical estimates derived from \cite{cuerno95,brown_prl,brown_prb}.

In summary, we have obtained a dynamical transition in the evolution of anisotropic patterns that illustrates the rich
phenomena that occur when taking into account the full anisotropy in this class of non-equilibrium systems. The transition merges
together two apparently opposed phenomena, like the selection of a typical wavelength and strong morphological fluctuations
leading to renormalization and scale invariance. Actually, the dominance of the latter at intermediate times
seem to be a requirement for the development of the pattern that later emerges. A dynamical role seems to be
played also by approximate solutions of the equation, such as cancellation modes. This suggest the interest of exploring
such type of solutions in other anisotropic, pattern forming systems. In its stabilized form, the isotropic KS equation
has been shown to provide a generic model for parity-symmetric systems featuring a bifurcation with a vanishing wave number
\cite{misbah94}. Thus, we expect the phenomenology of the aKS equation to apply quite generically. In particular, the transition
that we have discussed may offer an explanation for the recently observed ripple rotation in high-temperature IBS nanopatterning
experiments on Si surfaces \cite{brown_prl}. Further theoretical and experimental work is needed in order to elucidate the
degree to which this is actually the case, and the appearance of related phenomena in other pattern forming systems.

A.\ K.\ acknowledges financial support from the Alexander von Humboldt foundation.
M.\ N.\ and R.\ C.\ acknowledge partial support by MICINN (Spain) Grant No.\ FIS2009-12964-C05-01.

\bibliography{keller_bib}

\end{document}